# The ROSEBUD experiment at Canfranc : 2001 report

S. Cebrián[a], N. Coron[b], G. Dambier[b], E. García[a], I.G. Irastorza[a], J. Leblanc[b], P. de Marcillac[b]†, A. Morales[a], J. Morales[a], A. Ortiz de Solórzano[a], J. Puimedón[a], M.L. Sarsa[a], J.A. Villar[a]

a Laboratorio de Física Nuclear y Altas Energías, Univ. de Zaragoza, 50009 Zaragoza, Spain
b Institut d'Astrophysique Spatiale, CNRS, Bât. 121, 91405 Orsay Cedex, France

The ROSEBUD experiment for Direct Dark Matter detection settled in 1999 in the Canfranc Underground Laboratory. The first phase of the experiment was dedicated to the understanding and reduction of the radioactive background following successive removals of the radioimpure materials. Sapphire (25g, 50g) and germanium (67g) absorbers were used. Thresholds respectively lower than 1keV and 450 eV were achieved on these detectors. The second phase of the experiment plans to use scintillating bolometers to discriminate between recoiling nuclei and electrons. Prototypes using commercial $CaWO_4$ (54g) and BGO (46g) were designed for this purpose. While internal contamination was found and identified in both targets, neutron calibrations revealed their high discrimination power. A 6 keV threshold on the heat channel of the BGO bolometer points out the interest of such a novel material, for Dark Matter research on neutralinos having spin-dependent or spin-independent interactions.

## 1. INTRODUCTION

Among all detectors available for the direct detection of Dark Matter WIMPs through their scattering off nuclei, cooled massive bolometers present many advantages: very low threshold, good energy resolution, high conversion factor of incident energy into heat and a wide choice of absorbers. Moreover, an independent measurement of the ionisation power of the particles through charge or light collection may provide powerful discriminating tools against radioactive background.

The ROSEBUD experiment settled in 1999 in the Canfranc Underground Laboratory. As a first step, a small size, low cost, experiment was designed to establish the best operating conditions for Dark Matter detection. The same commercial transportable dilution refrigerator having a base temperature of 20mK was used at Orsay and Canfranc for, respectively, detector tests and observations in low background conditions.

† corresponding author
This work has been supported by the EU Network contract ERB FMRX-CT98-0167, by the French CNRS R&D program BOLERO, and the Spanish CICYT (contract AEN99-1033)

Sapphire bolometers allowed us to continuously improve the background while measuring it down to the 1keV level. Though, it appeared difficult to identify the remaining contaminant sources due to the low peak efficiency in such a low Z material [1].

Prototype bolometers more suited to deal with the background problem, either because its better identification of lines (Ge bolometer) or because their strong discrimination power ($CaWO_4$ and BGO scintillating bolometers) were recently made. The results of the tests with these prototypes are summarised hereafter.

## 2. A 67g GERMANIUM BOLOMETER

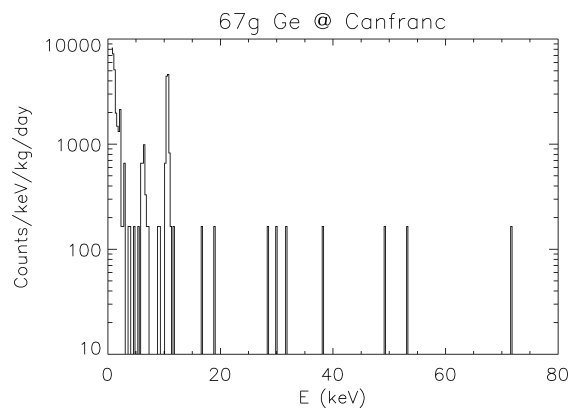

Figure 1

The background spectrum recorded at Canfranc by a 67g Ge bolometer (∅= h = 25mm) during a night is shown on Figure 1.

The average background level of 5 evts/keV/kg/day above 15keV is similar to that found on sapphire bolometers, indicating that they were seeing probably the same background. Note the very low threshold of this detector (420 eV at 5σ), the detection of cosmogenic $^{71}$Ge, as well as the steep increase of the spectrum at low energy, associated to non-anomalous events neither in pulse shape nor in rate.

## 3. SCINTILLATING BOLOMETERS

### 3.1. Common Set-up

Scintillating bolometers at 20mK were mounted according to the set-up shown in Figure 2.

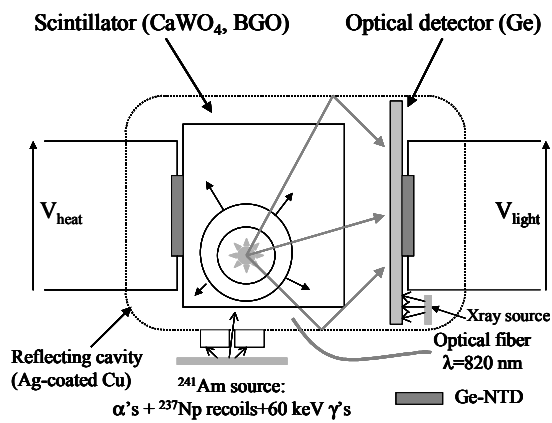

Figure 2

The absorber of the optical bolometer is a thin Ge disk (∅=25mm; thickness=100μm), looking at the scintillating crystal in an Ag-coated reflecting cavity. A backing $^{55}$Fe X-ray source allows to estimate its responsivity and the photon collection yield. Light pulses (λ=820nm) are sent regularly through plastic optical fibers to correct for any drift in both detectors. Temperatures are monitored by Ge-NTD sensors.

### 3.2. Study of a 54 g CaWO4 bolometer

A 54g $CaWO_4$ crystal [2], tested under neutron irradiation at Orsay, revealed relative light/heat amplitude ratios of 10 | 2.5 | 1 for, respectively, gammas | alphas | recoiling nuclei. Since a check of radioactivity on a low background n-type Ge spectrometer did not show any contamination, the $CaWO_4$ bolometer was then studied at Canfranc, where the discrimination plot of Figure 3 was recorded (15 hours). Threshold was about 45keV on

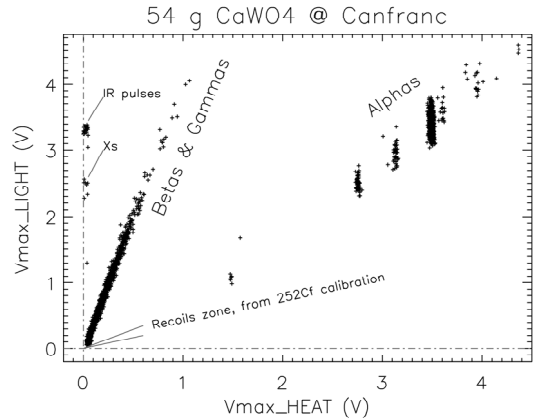

Figure 3

the heat channel, mainly due to microphonics

No recoil was registered in the "neutron" zone, as expected, but a high contamination in betas/gammas and alphas was clearly seen. The associated alpha spectrum is shown in Figure 4.

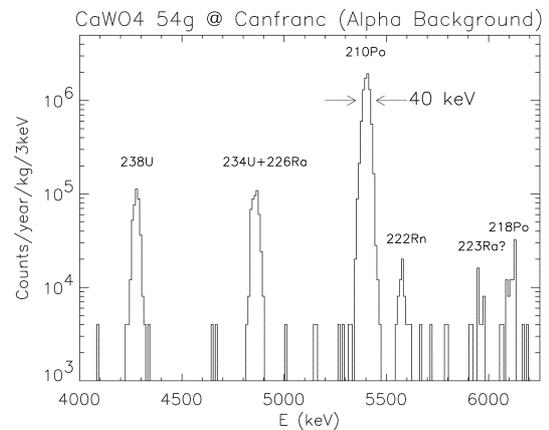

Figure 4

Alpha lines belong to the $^{238}$U chain. Further calibration with a $^{241}$Am source showed that the full energy $Q_\alpha$ of the alpha decays was recorded, which corresponds to an internal contamination. A very similar spectrum was reported on $TeO_2$ bolometers by the MIBETA experiment [3], but at a much lower level (by a factor $10^{-3}$) and was attributed to surface events. The continuum from 50 keV to about 1 MeV can be attributed to a contamination from $^{210}$Bi. The associated activity shows that $^{210}$Pb and its daughters are close to equilibrium (0.7 Bq/kg).

### 3.3. Study of a 46g BGO bolometer

A 46g BGO ($Bi_4Ge_3O_{12}$) bolometer was mounted [4], using the same optical bolometer as for $CaWO_4$, and studied at Orsay. The high discrimination power of this detector is clearly seen from Figure 5 ($^{252}$Cf neutron irradiation) and Figure 6 (background).

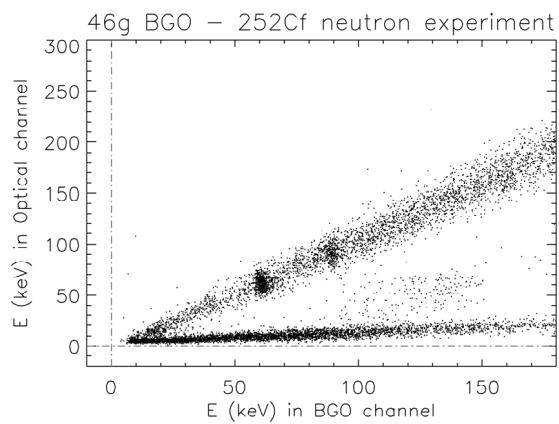

Figure 5

The light/heat factors and the collected light yields ($\approx 0.8 \%$ for gammas) are found very similar to those of $CaWO_4$, while the threshold is much lower for the BGO bolometer, bounded to 6 keV (5$\sigma$) due to the electronics.

An important $^{207}$Bi contamination, currently seen in all BGO detectors, was also identified here. It explains the 88keV line seen in the diagrams, filled by the X rays and electron cascades from the K-shell of $^{207}$Pb. A $^{207}$Bi activity of about 3Bq/kg could be deduced from this line. The 60keV gammas from a $^{241}$Am calibration source are also detected ($\Delta E_{FWHM} < 4$ keV). As the alpha source was unprotected, the BGO bolometer registered as well the recoiling $^{237}$Np nuclei ($E_{max} \approx 93$ keV). The light emission efficiency of these events is found closed to that of the nuclei (mainly O) recoiling after neutron scattering. The hits in the intermediate zone between the gammas and the recoiling atoms are well explained by the BGO bolometer ability of summing $^{237}$Np recoils and 60keV gamma paired events that pass the collimator in coincidence.

Recoils from alpha sources, such as $^{206}$Pb from $^{210}$Po, might reveal very useful – and cheep - tools in the study of energy conversion efficiency of recoils,

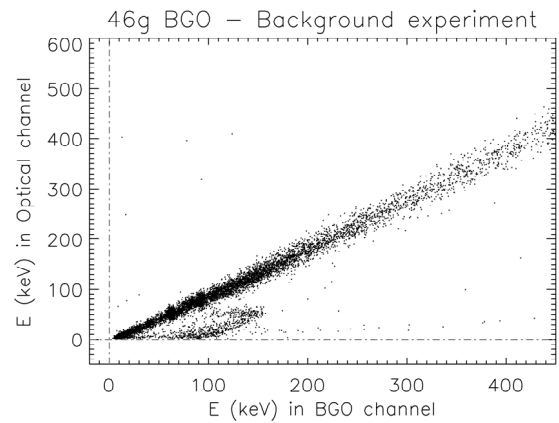

Figure 6

since they transfer easily their energy to heavy nuclei (as Bi in BGO) through collision cascades.

### 4. CONCLUSION AND PROSPECTS

New performant bolometers have been designed for ROSEBUD that could help to better understand the background. Scintillating prototypes from unselected commercial samples of $CaWO_4$ and BGO have been tested, showing internal radioactive contamination. Nevertheless, their high discriminating power may render them usable in a long duration experiment. But since we plan to run together as many different targets as possible, our first goal is now to find clean scintillating crystals that will not contaminate the other "single channel" detectors (sapphire,…).

The good threshold of the 46g BGO bolometer makes it a very attractive candidate as a target for a Dark Matter experiment, thanks to its unique high spin content (natural Bi is 100% $^{209}$Bi ; J=9/2).

### 5. REFERENCES

1. S. Cebrián et al., IDM2000 proceedings, 361-6
2. *PI-KEM, Shropshire, SY4 5HE, UK (2000)*
3. E. Fiorini, personal communication
4. *Crismatec, 38610 Gieres, France (1997)*